\documentclass[twocolumn]{aastex62}
\usepackage{amsmath}
\usepackage{mathrsfs}
\usepackage{multirow}
\usepackage{lineno}
\usepackage{tabularx}
\usepackage{subfigure}
\usepackage[normalem]{ulem}
\usepackage{graphicx}
\usepackage{booktabs}
\usepackage{url}
\usepackage{color,subfigure,lineno}
\usepackage[T1]{fontenc}
\usepackage{academicons}
\usepackage{xcolor}

\usepackage{hyperref}
\usepackage{orcidlink}

\usepackage{hyperref}

\newcommand{\comments}[1]{}

\newcommand{\halpha}{H{$\alpha$}}
\newcommand{\solarm}{M{$_{\odot}$}}
\newcommand{\landa}{$\lambda$}

\newcommand*{\rom}[1]
{\expandafter\@slowromancap\romannumeral #1@}

\newcommand{\snu}{\affil{Department of Physics \& Astronomy, Seoul National University, Seoul 08826, Republic of Korea, jhwoo@snu.ac.kr, ayubinia\_84@snu.ac.kr}}
\newcommand{\mc}{\affil{McWilliams Center for Cosmology, Department of Physics, Carnegie Mellon University, Pittsburgh, PA 15213}}
\newcommand{\unist}{\affil{Artificial Intelligence Graduate School, UNIST, Ulsan, Republic of Korea}}

\begin{document}

\title{Prediction of Star Formation Rates Using an Artificial Neural Network}

\author{Ashraf Ayubinia\orcidlink{0009-0002-9526-5820}}\snu
\author{Jong-hak Woo\orcidlink{0000-0002-8055-5465}}\snu
\author[0000-0000-0000-0000]{Fatemeh Hafezianzadeh} \mc
\author[0000-0000-0000-0000]{Taehwan Kim}\unist
\author{Changseok Kim\orcidlink{0000-0002-2156-4994}}\snu

\begin{abstract}
In this study we develop an artificial neural network to estimate the infrared (IR) luminosity and star formation rates (SFR) of galaxies. Our network is trained using 'true' IR luminosity values derived from modeling the IR spectral energy distributions (SEDs) of FIR-detected galaxies. We explore five different sets of input features, each incorporating optical, mid-infrared (MIR), near-infrared (NIR), ultraviolet (UV), and emission line data, along with spectroscopic redshifts and uncertainties. All feature sets yield similar IR luminosity predictions, but including all photometric data leads to slightly improved performance. This suggests that comprehensive photometric information enhances the accuracy of our predictions. Our network is applied to a sample of SDSS galaxies defined as unseen data, and the results are compared with three published catalogs of SFRs. Overall, our network demonstrates excellent performance for star-forming galaxies, while we observe discrepancies in composite and AGN samples. These inconsistencies may stem from uncertainties inherent in the compared catalogs or potential limitations in the performance of our network.
\end{abstract}


\section{Introduction}
The star formation rate (SFR) is a fundamental parameter in galaxy evolution. The star formation process releases energy and new chemical elements that directly impact the evolution of galaxies. Therefore, accurate estimation of SFR is crucial for unraveling the intricate physics underlying galaxy evolution and cosmic star formation history.  

A common method of quantifying the SFR of galaxies relies on observations of SFR indicators \citep[]{Moustakas06, Wuyts11, Schreiber20}.
In particular, the ultraviolet (UV) continuum emission, which is predominantly emitted by the young stellar populations (which typically fall into the O and B spectral classes), can provide a powerful tool for assessing star formation. Given the short-lived nature of massive stars, the UV continuum is associated with the ongoing star formation activity \citep[$\sim$ 100 Myr;][]{Salim07,Buat08,Davies16}. However, employing UV as a diagnostic tool presents several challenges. In addition to the heavy absorption of UV emission by dust \citep[e.g.,][]{{Calzetti94}}, the relationship between UV luminosity and SFR is influenced by gas metallicity \citep[][]{{Madau14}}. Galaxies with higher metallicity emit less UV radiation per unit star formation compared to those with lower metallicity, introducing systematic uncertainty in SFR estimates. Furthermore, the conversion factor {from} UV luminosity to SFR is not universally fixed, as it can vary depending on {various factors i.e., the stellar initial mass function (IMF)} \citep[e.g., ][]{{Raiter10}}. These factors contribute to uncertainties in the estimation of SFR \citep[]{Kennicutt98}.

With the advent of large spectroscopic surveys, nebular emission-line SFR tracers have been extensively applied to both local and distant galaxies \citep[e.g., ][]{Doherty06, Villar08}. The \halpha~emission arises from the recombination of interstellar gas ionized by most massive stars ($\gtrsim 15$ \solarm). Hence, \halpha-based SFR represents the instantaneous measure of the SFR over the past $\sim$ 3-10 Myr. The stability of the \halpha~line strength against density or temperature variations makes it one of the most reliable tracer of star formation \citep[]{Kennicutt98}. {As the \halpha~emission line moves out of the optical window into the near-infrared (NIR) for objects at high redshifts ($z \gtrsim 0.4$), the [O II] \landa3727 emission line is frequently adopted as an alternative SFR indicator} \citep[e.g., ][]{Kennicutt92, Rosa02, Argence09, Gilbank10}.

{Dust in the ISM introduces additional uncertainty, absorbs UV/optical light, and re-emits it in the form of infrared (IR) radiation.}
Different physical processes dominate different parts of the IR Spectral Energy Distribution (SED) of galaxies. The diffuse emission from polycyclic aromatic hydrocarbon (PAH) molecules dominates in the $\lambda\sim$ 3--20$\mu $m \citep[]{O'Dowd09, Calzetti13}. At longer wavelengths, the warm dust continuum, {dominated by very small dust grains are stochastically heated by the diffuse interstellar radiation field at high temperatures and prevails in the $\lambda\sim$ 20--60 $\rm \mu $m domain \citep[]{Calzetti13}.}
Beyond $\sim$ 60 $\mu$m, the cold dust continuum is produced by big dust grains in thermal equilibrium at cool temperatures \citep[about 15-20 K or above,][]{Draine03, Barroso13} dominates. This component is responsible for the bulk of the far-infrared (FIR) emission where the SED peaks. Since the dust absorption cross-section peaks in the UV, the FIR emission becomes a highly sensitive tracer of the young stellar population and SFR. Despite its advantages, the limited sensitivity and spatial resolution of FIR observation facilities have constrained the application of FIR for SFR diagnostics {\citep[e.g.,][]{Takeuchi2010, Matsuoka15}.}

Along with massive volumes of astronomical data, Machine learning approaches have contributed greatly to the advancement of astronomy \citep[see the reviews by][]{Baron19, Fluke20, Haghighi22, Webb23}. In addition to the classification \citep[][and references therein]{Kuntzer16, Clarke20, Zeraatgari24a}, machine learning is widely applied to regression problems, enabling {precise predictions} of quantities such as SFR. For example, using a sample of 1136 galaxies in the Herschel-SDSS Strip 82, \cite{Ellison16} trained an artificial neural network (ANN) and predicted the total IR luminosity estimates for $\sim$ 330,000 SDSS galaxies. Similarly, \cite{Bonjean19} employed Random Forest to estimate SFRs and stellar masses for galaxies at  $0.01 < z < 0.3$. Additionally, \cite{Surana20} leveraged deep learning techniques on data from the Galaxy And Mass Assembly (GAMA) survey to estimate stellar masses, SFRs, and dust luminosity. These studies \citep[see also][]{Dobbels20, Narendra22, Euclid23, Hunt24, Zeraatgari24b, Gai24} collectively underscore the effectiveness of machine learning in estimating SFRs and related parameters across diverse datasets and methodologies, thus advancing our understanding of galaxy evolution and cosmology. 

{While SFR estimations based on narrower bands, such as NIR \citep[e.g.,][]{Neufeld24}, mid-infrared \citep[MIR, e.g.,][]{Sureshkumar23}, and FIR \citep[e.g.,][]{Picouet23}, are available, in this study, we employ an artificial neural network on multi-band data to estimate total IR luminosity as a key indicator of SFR.}
By utilizing solely photometric data and incorporating spectroscopic redshift of galaxies, {our methodology not only enables the prediction of IR luminosity and SFR for a large population of galaxies but also provides reliable and accurate estimates of SFRs.} In Section \ref{data}, we explain the data collection process for estimating true IR luminosities and training our network. Section \ref{network} illustrates the structure of our network and the training process. The results are presented in Section \ref{results}, {and we summarize} the application of our neural network on unseen data in Section \ref{discussion}, and we {present} a summary in Section \ref{summary}.

\section{Data}\label{data}
As we plan to implement our network on the Sloan Digital Sky Survey galaxies, we construct our training sample based on SDSS DR17 \citep[]{Abdurro2022}. {We integrate data from the $photoObj$ catalog, which includes photometric data from previous data releases, with spectroscopic data from the $specObj$ catalog. This latter catalog incorporates new spectra obtained from the SDSS component of the extended Baryon Oscillation Spectroscopic Survey (eBOSS) for all targets classified as 'GALAXY' or 'QSO'.} The spectroscopic survey observes galaxies with a Petrosian magnitude of $r_{Petro} < 17.77$ for the main galaxy sample (Strauss et al. 2002) and $r_{Petro} < 19.5$ for the luminous red galaxy (LRG) sample \citep[]{Eisenstein2001}. We merged the two catalogs by aligning the OBJID in $photoObj$ with the 'BESTOBJID' in $specObj$. Only targets with reliable redshift measurements are included (ZWARNING = 0). This selection comprises $\sim$ 1.2 million galaxies, which we refer to as our parent sample. After assembling our parent sample, we gather two sets of data. Initially, we acquire FIR and MIR data to model the IR SED of the training sample, aiming to {determine} the true values of IR luminosity. Subsequently, we collect additional data to serve as input features for our network, as outlined below.\

\subsection{{IR} Data Collection for SED Fitting}
\begin{figure}[htbp]
\centering
\includegraphics[width=0.4\textwidth]{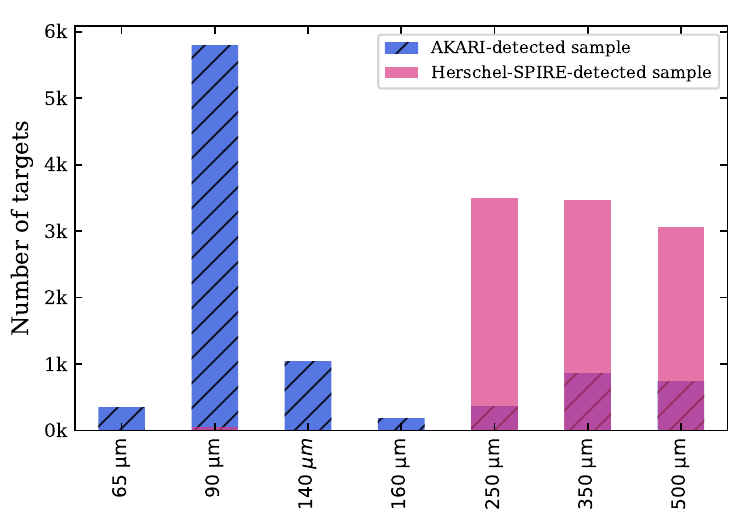}
\caption{Distribution of FIR-detected galaxies detected by AKARI (blue) and Herschel-SPIRE (red) across various FIR bands.}
\label{fig:FIR_data_points}
\end{figure}

The goal of the present study is to predict SFR using IR luminosity. The FIR photometric measurements are crucial to identify IR galaxies whose SEDs peak around 100 $\rm \mu m$. The AKARI FIR all-sky survey carried out by using Far-Infrared Surveyor \citep[FIS;][]{Kawada2007} with superior sensitivity, spatial resolution \citep[$\rm FWHM = 1^{'}-1.5^{'}$,][]{Doi2015} and wider wavelength coverage ($\rm 50-180\ \mu m$) than Infrared Astronomical Satellite (IRAS), {providing} a unique and complementary perspective on the universe. We cross-match our parent sample with the AKARI/FIS bright source catalog version 2 \citep[]{Yamamura2018}, which includes FIR flux measurements for 918,056 IR sources with a matching radius $12^{''}$, corresponding to 2$\sigma$ positional uncertainty of AKARI. Our target selection requires exceeding the $5\sigma$ detection limit in at least one band, which translates to 2.4, 0.55, 1.4, 6.3 Jy for the 65, 90, 140, 160 $\rm \mu m$ bands \citep[]{Kawada2007}, respectively. {We obtain 5856 matched targets, of which $\sim$ 15\% are detected in at least one band of the Herschel SPIRE survey.} In addition to AKARI data, we also utilize the Herschel FIR data to increase the size of FIR-detected galaxies. To do this, we cross-match our parent sample with the Herschel Stripe 82 Survey \citep[HerS,][]{Viero14}. HerS covers approximately half of the 150 $\rm deg^{2}$ of the deep SDSS Stripe, providing a 250 $\rm \mu m$-selected catalog of 3.3 $\times 10^{4}$ sources detected at $> 3 \sigma$ (including confusion noise), {where the completeness is estimated to be 50\% and the false detection rate to be less than 1\% \citep[]{Viero14}.} We find that 3540 galaxies from our parent sample are included in this catalog, of which $\sim$ 3\% are detected in at least one band of the AKARI survey. We eliminate 48 targets already included in the AKARI-detected sample. Figure \ref{fig:FIR_data_points} illustrates the distribution of 9348 FIR-detected galaxies across various FIR bands.\

The MIR data of our FIR-detected galaxies are collected from Wide-field Infrared Survey Explorer \citep[WISE,][]{Wright10} {which surveyed} the entire sky in four infrared wavelengths of 3.4, 4.6, 12, and 22 $\rm \mu m$ (hereafter W1, W2, W3, and W4). We eliminate sources flagged as either spurious (created by diffraction spikes or astrophysical sources with photometry contaminated by diffraction spikes) or as contaminated by moonlight in the AllWISE catalog. Additionally, some objects with null values for flux uncertainties are removed from the sample. It results in 7162 IR galaxies suitable for fitting their IR SEDs.

\subsubsection{Estimation of True IR Luminosity}
To obtain true values of IR luminosity for our training sample, we fit the IR component of the SEDs using IR flux measurements from the AKARI and Herschel surveys. When IRAS data are available, we include them to further improve the accuracy and quality of our fits. In this process, various IR templates are employed through CIGALE fitting to accurately model the SEDs.

In CIGALE, there are different IR templates \citep[]{Dale2014, Draine07, Draine14}. In this study, we estimate dust luminosity using the model described in \cite{Casey12}. 
This model integrates a modified blackbody with a MIR continuum power law, effectively addressing both large-scale cold dust emission and warm dust emission. {We note that while this fitting technique does not account for PAH emission in the MIR, as its influence on integrated FIR properties is negligible \citep[]{Casey12}, it has been shown to perform well in both the case of sources with or without active galactic nuclei (AGN) \citep[]{Shimizu15}. }

To ensure consistency in our SED fitting results, we use two additional dust templates. First, we fit the IR SEDs using the Chary-Elbaz IR template library \cite[]{Chary01}, which includes 105 templates based on data from the IR Space Observatory  \citep[ISO;][]{Kessler96}, IRAS, and SCUBA. These templates cover a wide range of IR luminosities from normal star-forming galaxies to ultra-luminous infrared galaxies (ULIRGs) but do not account for AGN contributions, so we exclude W1, W2, and W3 data. Second, we use the models from \cite{Draine07} and \cite{Draine14}, which feature a dust mixture of amorphous silicate, graphite grains, and PAHs. These models distinguish between diffuse and star-forming dust components, with the 2014 update offering a variable power-law index and expanded radiation field ranges for more accurate and flexible fitting of IR data.

\begin{figure*}[htbp]
\centering
\includegraphics[width=0.9\textwidth]{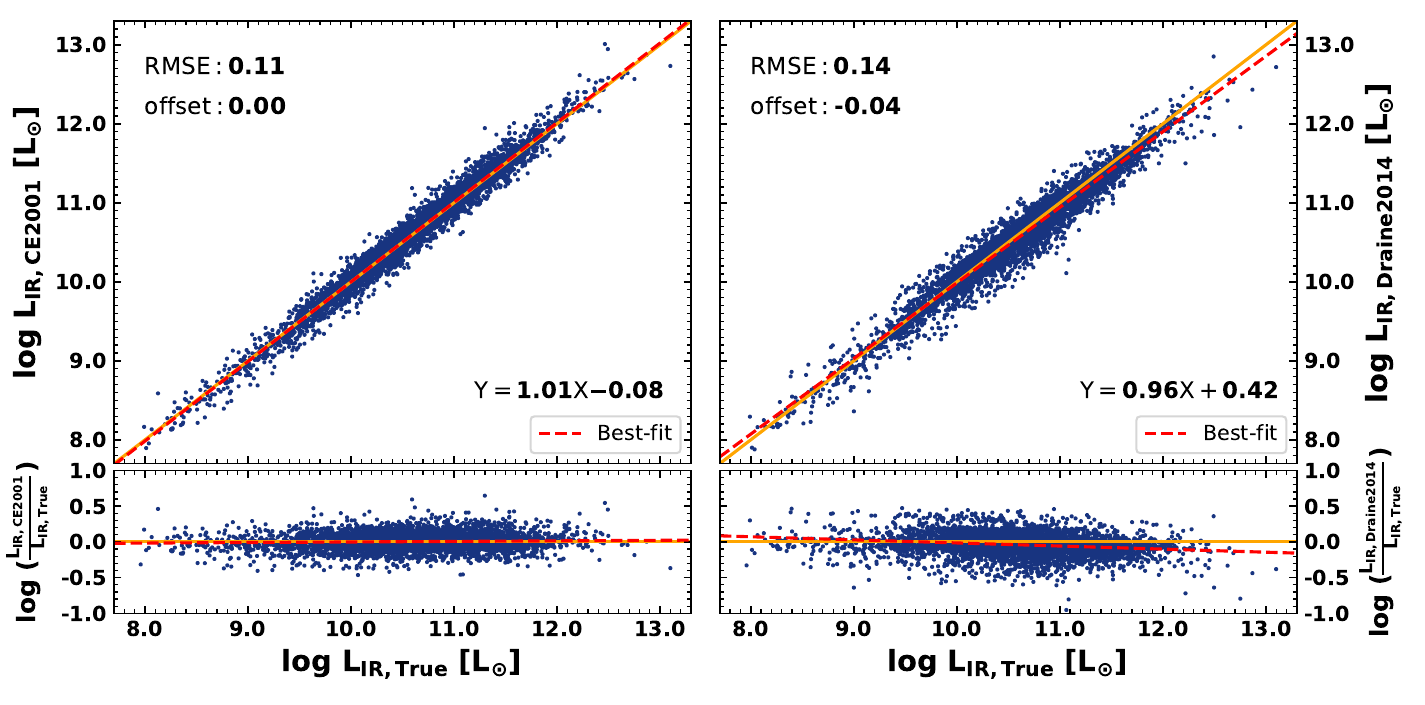}
\caption{Comparison between true IR luminosity estimated using the model presented in \cite{Casey12} with those estimated from templates described in \cite{Chary01} (left panel) and \cite{Draine14} (right panel). The red dashed line shows one-to-one correlation. {The orange line represents the one-to-one correlation, and the red dashed line indicates the best-fit. The equation of the best-fit is displayed on each plot.}}
\label{fig:true_val}
\end{figure*}

Figure \ref{fig:true_val} compares IR luminosity estimates derived from three different sets of templates. {The small scatter, indicated by the low Root Mean Square Error (RMSE), which quantifies the average magnitude of deviations between predicted values and actual outcomes, along with the negligible offset between estimates, indicates minimal deviations. This demonstrates the consistency in IR luminosity estimates and suggests they are not strongly dependent on the choice of IR templates.} Hence, we can proceed confidently with the results obtained from the template provided by \cite{Casey12} using CIGALE \citep[]{Boquien19, Yang20}.

\subsection{Data Collection for Training Network}
In the realm of neural networks, the term ``features'' refers to the input variables that neural networks process during the training phase to determine patterns, relationships, and representations within the data. In our study, we present five sets of features, evaluating the performance of the network in each case. Motivated by previous studies attempting to estimate SFR using machine learning techniques \citep[]{Ellison16, Bonjean19, Surana20}, our initial training involves the primary feature set, encompassing optical and MIR photometric data, along with SDSS redshift, and their corresponding errors. {Despite training our network with photometric redshift as well, the results showed minimal change, reinforcing our decision to prioritize spectroscopic redshift due to its higher accuracy.}

\begin{figure*}
  \centering
  \subfigure{%
    \includegraphics[width=0.79\textwidth]{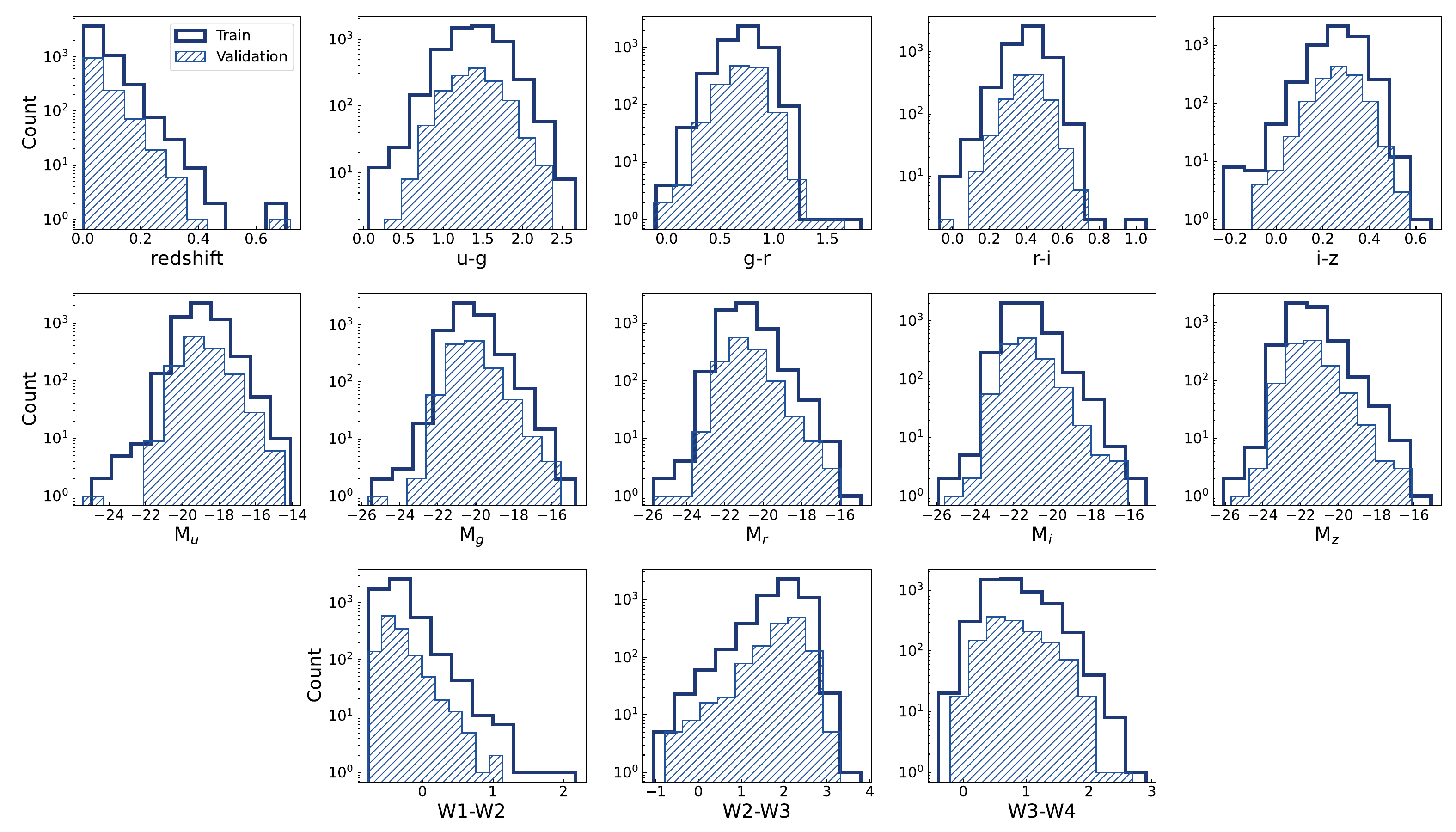}%
    }\hspace{0.cm}
    \subfigure{%
    \includegraphics[width=0.41\textwidth]{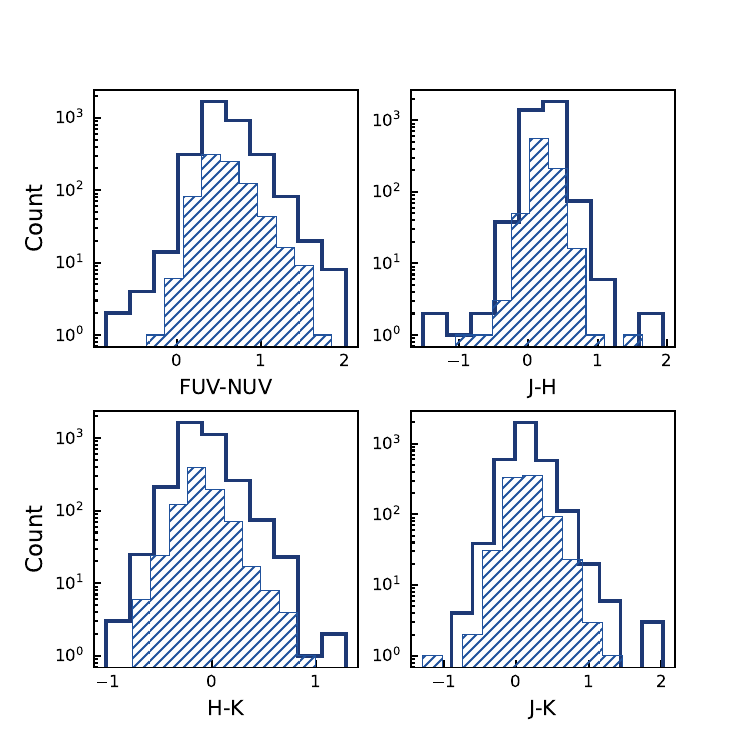}%
  }%
    \subfigure{%
    \includegraphics[width=0.41\textwidth]{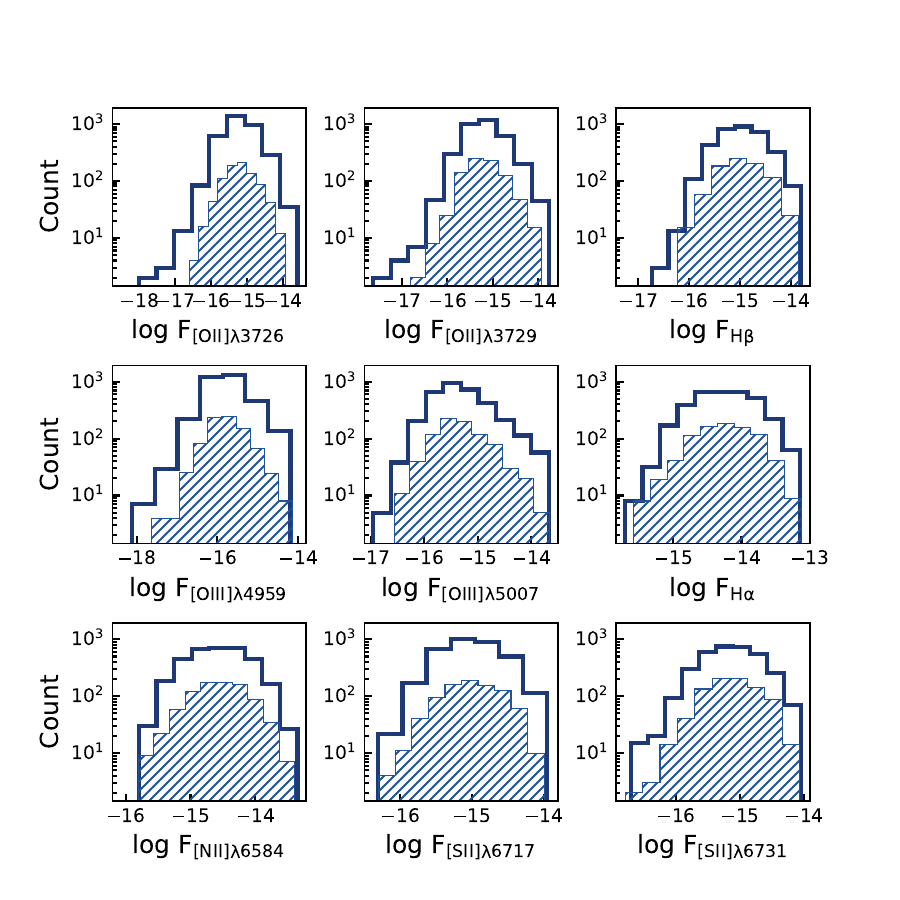}%
  }%
  \caption{The distribution of the primary (top), NIR and UV (bottom left), and emission line (bottom right) features for {training data (solid histograms) and the validation data (dashed histograms)}. Magnitudes are in AB system and emission line flux unit is $\rm erg\ cm^{-2}\ s^{-1}$.}
  \label{fig:feature_dist}
\end{figure*}

Subsequently, we incorporate additional features including NIR and UV {photometry} as well as emission lines in a stepwise manner, each time retraining the network with the augmented data (see Table \ref{tab:features}). We collect NIR and UV photometric data from Two Micron All Sky Survey  \citep[2MASS,][]{Huchra05, Huchra12} and {Galaxy Evolution Explorer \citep[GALEX,][]{Martin05, Bianchi14}} using a searching radius $12^{''}$. Emission lines flux are taken from MPA-JHU catalog \citep[]{Abazajian09}. To ensure robustness, we exclude targets with measurements carrying significant uncertainties exceeding $\rm 3\sigma$, thereby refining the dataset for more reliable model training. 
As a result, we have a sample of 6426 FIR-detected sources with the primary feature set as input data. This sample is divided into training and validation sets, with 80\% allocated for training and 20\% for validation (see Section \ref{network}). Figure  \ref{fig:feature_dist} shows the distribution of features for both the training and validation datasets.

\begin{table*}
\centering
\caption{List of feature sets.}
\begin{tabular}{|c|c|c|c|}
\hline
\multirow{1}{*}{Primary Features} & \multirow{1}{*}{NIR Features} & \multirow{1}{*}{UV Features} & \multirow{1}{*}{Emission Lines Features} \\
\hline\hline
$z$: redshift & J-H  & FUV-NUV  & luminosity of [O II] $\lambda 3727$ \AA~emission line \\
$\rm M_{u}$: absolute u-band magnitude & H-K &  & luminosity of [O II] $\lambda 3729$ \AA~ emission line\\
$\rm M_{g}$: absolute g-band magnitude & J-K &  & luminosity of [O III] $\lambda 4959$ \AA~ emission line \\
$\rm M_{r}$: absolute r-band magnitude & & & luminosity of [O III] $\lambda 5007$ \AA~emission line\\
$\rm M_{i}$: absolute i-band magnitude & & & luminosity of $\rm H\alpha$ $\lambda 6563$ \AA~emission line\\
$\rm M_{z}$: absolute z-band magnitude & & & luminosity of $\rm H\beta$ $\lambda 4861$ \AA~emission line \\
u - g& & & luminosity of [N II] $\lambda 6582$ \AA~emission line\\
g - r & & & luminosity of [N II] $\lambda 6582$ \AA~emission line\\
r - i & & & luminosity of [S II] $\lambda 6717$ \AA~emission line\\
i - z & & & luminosity of [S II] $\lambda 6731$ \AA~emission line\\
w1-w2 & & & \\
w2-w3 & & & \\
w3-w4 & & & \\
\hline
\end{tabular}
\\
Column1: The primary features, Column 2 to 4: NIR colors, UV colors and emission lines, respectively. We incorporated additional features including NIR and UV photometric colors as well as emission lines in a stepwise manner, each time retraining the network with the augmented data.
\label{tab:features}
\end{table*}

\section{Artificial neural network model}\label{network}
{Our objective is to design a network capable of predicting IR luminosity accurately.} After exploring various machine learning models {such as XGBoost and Random forest, we find that a simple artificial neural network (ANN) exhibits the most promising performance on our data. {Artificial Neural Networks \citep[][]{Uhrig95,Abraham05} are computational models inspired by the structure and function of the human brain. They are highly effective at identifying patterns and revealing intricate relationships in data, primarily serving three key functions; regression (as shown in this study), classification, and control/optimization.} These networks are built using layers of interconnected processing units called artificial neurons. Each neuron receives information, performs calculations, and transmits the results to the next layer. A key strength of neural networks is their ability to model complex, non-linear relationships without needing pre-existing knowledge about the structure of data. However, building an effective neural network requires experimentation. This involves fine-tuning various parameters such as the number of neurons, activation functions, learning rate, training iterations (epochs), optimization algorithms, and regularization techniques to prevent overfitting.}

{Our model is designed as a multilayer perceptron (MLP) that comprises one input layer, one hidden layer, and one output layer.} The layer contains 128 neurons and takes the input directly, processing it to produce the output. This simple structure leverages the ability of the dense layer to model complex relationships in the data, even with a straightforward configuration. Each neuron uses the Rectified Linear Unit \citep[ReLU,][]{LeCun15, Agarap18} activation function, which performs a non-linear transformation on the output received from the previous layer. {Our regression model employs a linear activation function in the output layer, enabling continuous value prediction. Note that while this allows for potentially unbounded output, other model factors may influence the actual range.} Additionally, the He uniform initializer \citep[]{He16} is used to initialize the weights of the layers, and L1 and L2 regularization is applied to each layer to prevent overfitting by penalizing large weights.  

{The model is optimized using the mean squared error (MSE) loss function and the Adam optimizer, a stochastic gradient descent (SGD) variant. Adam adapts the learning rate by estimating first-order moments (the mean of the gradients) and second-order moments (the uncentered variance of the gradients). This allows Adam to perform well in various situations, especially when dealing with sparse gradients or noisy data, by adjusting the learning rate dynamically for each parameter during training \citep[]{Kingma15}. The initial learning rate is set to 0.001.} The model evaluates its performance using multiple metrics, including MSE, RMSE, and $\rm R^{2}$ score (that indicates the proportion of the variance in the dependent variable that is predictable from the independent variables), all of which are considered important for assessing the performance of the model. 
To effectively monitor the performance of our model during training and combat local minima in the loss function, we employ two key techniques. First, we dynamically adjust the learning rate by reducing it when the validation loss plateaus. Second, we implement early stopping, which halts training if the validation loss fails to improve over a specified number of epochs.
We split our input data (6246 FIR-detected sources) into training and validation sets, allocating 80\% for training and 20\% for validation in each run. Using the training data, the model is trained, and then its performance is evaluated using the validation data. Model refinement and enhancement are driven by the error calculated on the validation set, guiding iterative improvements.

 \section{Results}\label{results}
This section presents the results of all experiments conducted using the aforementioned architecture. Table \ref{tab:statasitics} shows the results for each model, including different sets of input features. {To ensure the robustness of our training process and avoid local minima, we trained each network 10 times. The differences between these iterations are negligible, suggesting that our models are robust. Therefore, we consider the best results from each network.}

Although all five feature sets exhibit identical overall performance, the output from the network shows slightly better statistical metrics {of both training and unseen data when all photometric features are utilized (particularly in the case of star-forming galaxies, see Table \ref{tab:statasitics})}. Hence, this particular model (i.e., Main Model in Table \ref{tab:statasitics}) {is preferable due to its better performance, which includes a higher  $\rm R^{2}$ score and lower MSE on both training and test sets.}

\begin{figure}[htbp]
\centering
\includegraphics[width=0.45\textwidth]{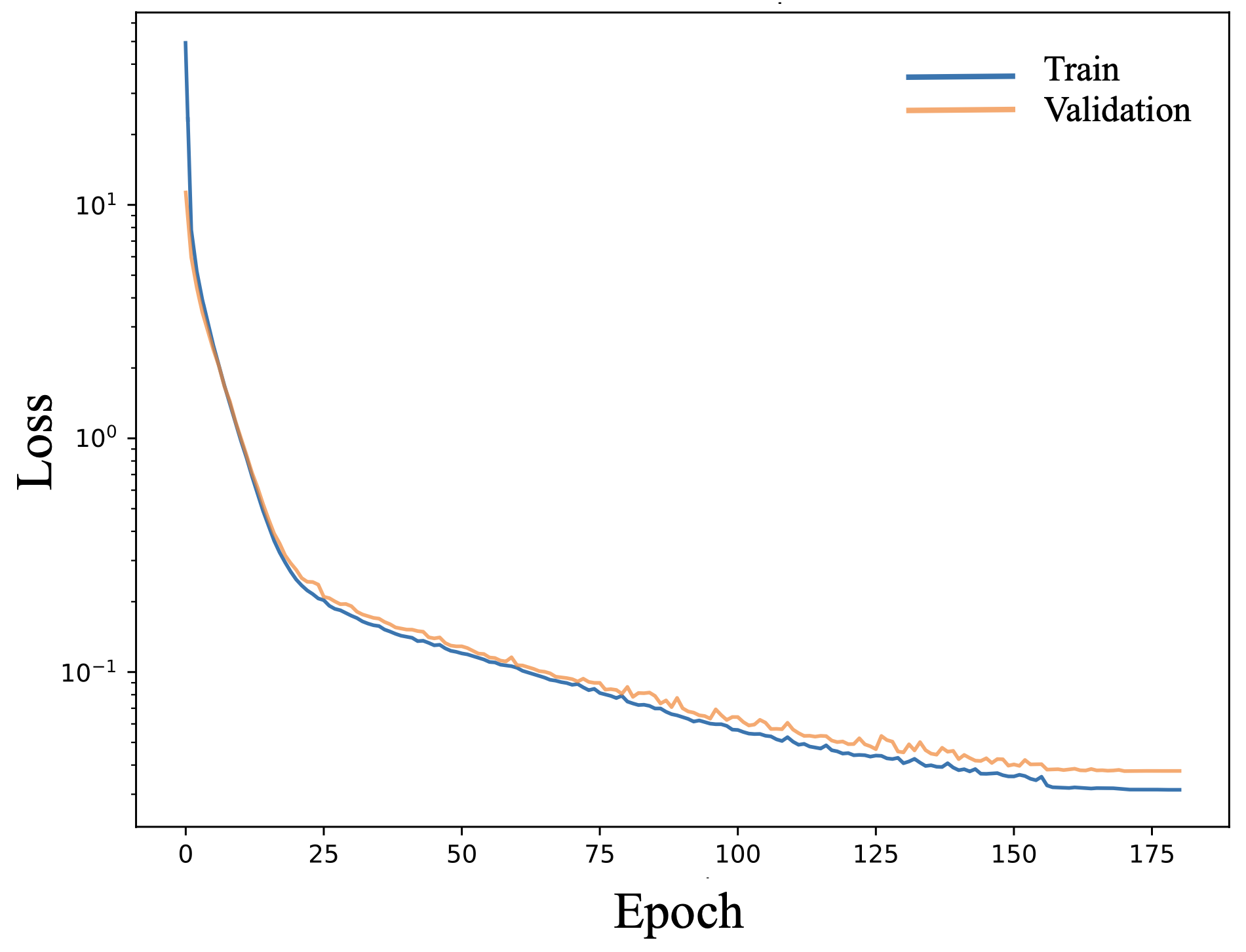}
\caption{The loss function for training (blue) and validation (orange) data as a function of epoch for the main model network. The gap between the loss function of the training and validation is negligible.}
\label{fig:loss}
\end{figure}

The loss function serves as a crucial metric that quantifies the disparity between the true values and the predictions generated by the network during the training process. During training, the network iteratively adjusts its parameters to minimize the loss function, effectively fine-tuning its predictive capabilities. A decreasing trend in the loss function over epochs indicates that the network is learning to better approximate the true values. Figure \ref{fig:loss} illustrates the loss function for our main model. In our training process, we employ a learning rate reduction strategy where the learning rate decreases by a factor of 0.1. This reduction factor is applied periodically during training epochs. By gradually decreasing the learning rate, we allow the optimization process to converge more effectively towards an optimal solution, potentially preventing overshooting of the minimum loss. This strategy often helps improve the stability and convergence of the training process, leading to better generalization performance on unseen data.

\begin{figure*}[]
\centering
\includegraphics[width=1\textwidth]{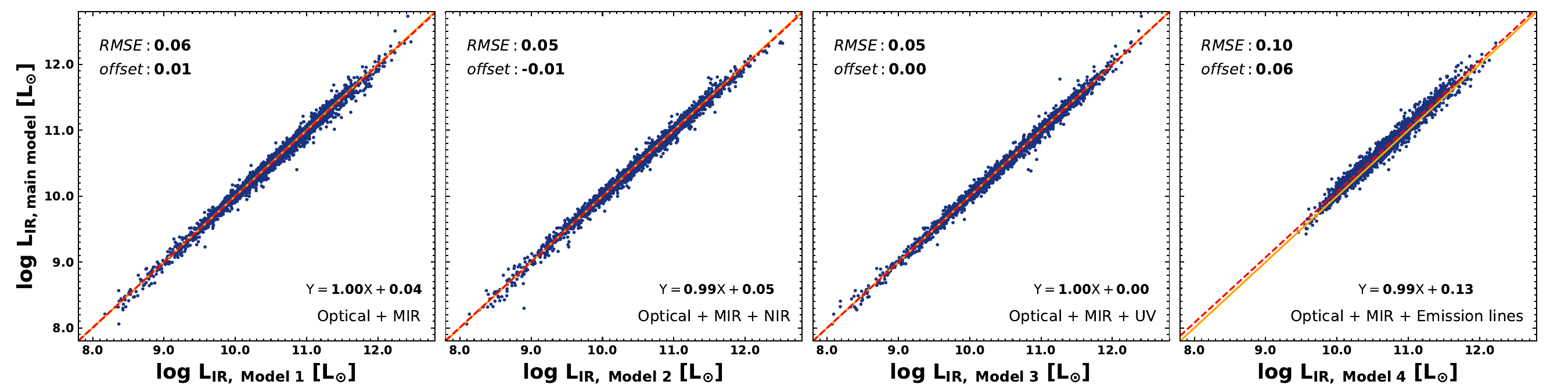}
\caption{Comparison between IR luminosity estimated using the mainT model (Y-axis) with those estimated from other models (X-axis); Model 1: Optical + MIR, Model 2: Optical + MIR + NIR, Model 3: Optical + MIR+ UV, Model 4: Optical + MIR + emission lines. {The orange line represents the one-to-one correlation, and the red dashed line indicates the best-fit. The equation of the best-fit is displayed on each plot.}}
\label{fig:comparisons}
\end{figure*}

{In Figure \ref{fig:comparisons}, outputs of four models are scrutinized against those of the main reference model, encompassing both training and validation datasets. It is clear that these models provide consistent results with marginal offsets and scatters. Following these comparisons, we proceed with our analysis using the Main Model.}

\begin{table*}[htbp]
\centering
\caption{The statistical results of our trained network using different feature sets.}
\includegraphics[width=0.8\textwidth]{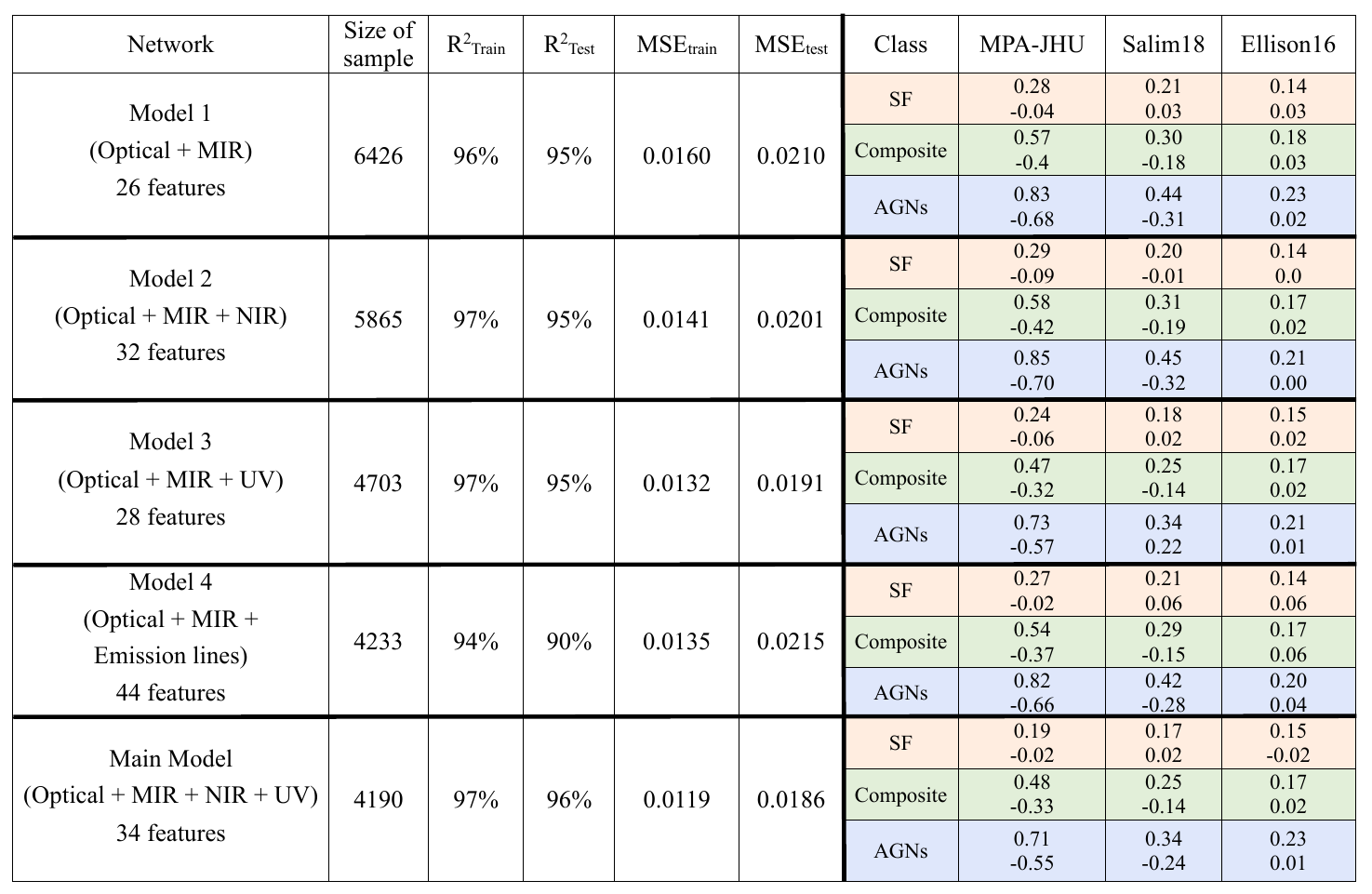}\\
\centering
\footnotesize Column 1: Name of model and the number of features, Column 2: Size of sample (training plus validation data), Column 3, 4: $\rm R^{2}$ score for training data and validation data, Column 5, 6:  Mean square error (MSE) for training data and validation data, Column 7: target classification based on MPA-JHU catalog, Columns 8, 9, 10: In each cell, the top number represents the RMSE, while the bottom number indicates the offset between the values provided by the catalogs and the predicted SFRs (see Figure \ref{fig:compared_catalogs}).
\label{tab:statasitics}
\centering
\end{table*}

\begin{figure*}
  \centering
  \subfigure{%
    \includegraphics[width=0.49\textwidth]{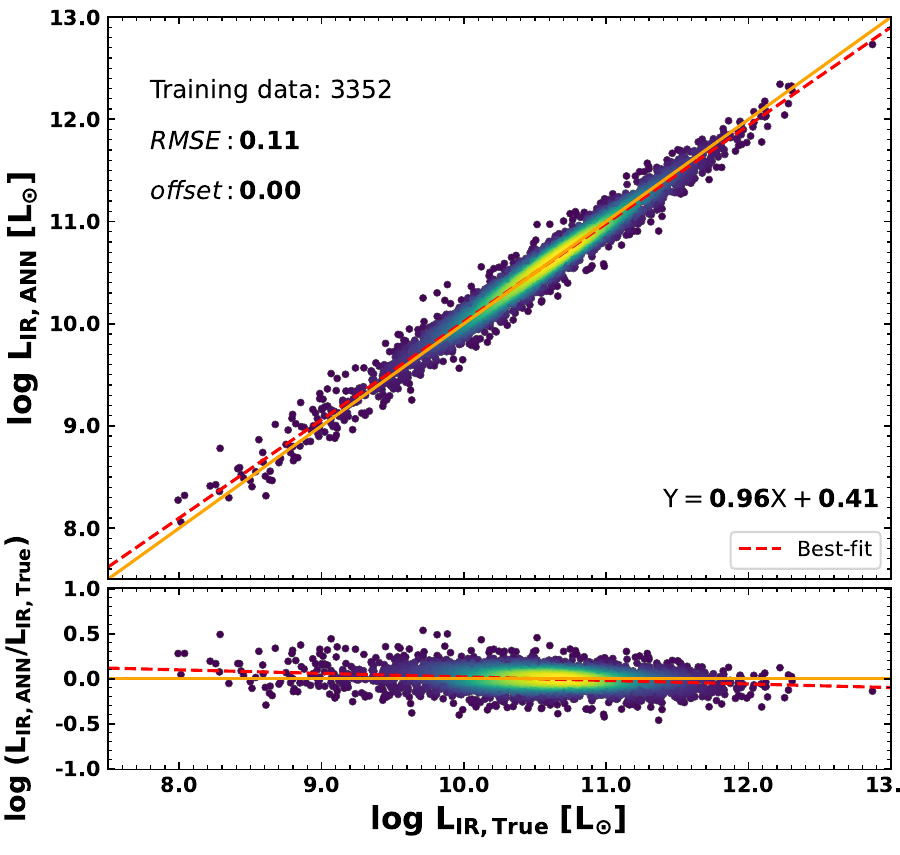}%
    }\hspace{0.2cm}
    \subfigure{%
    \includegraphics[width=0.49\textwidth]{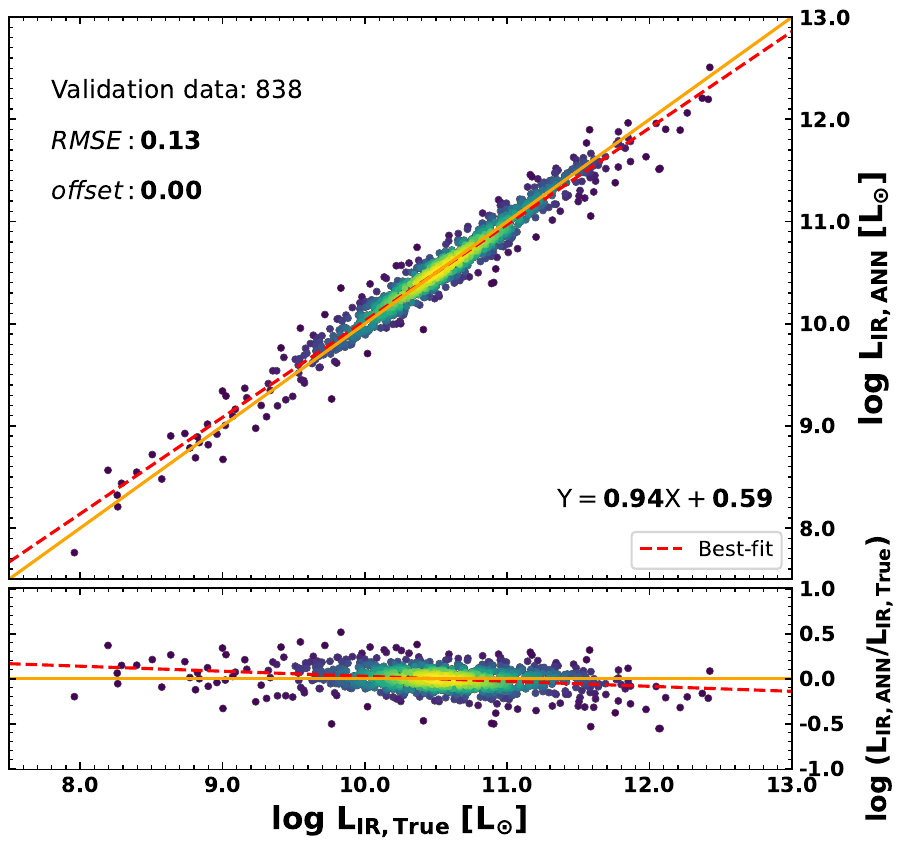}%
  }%
  \caption{The comparison between the true IR luminosities estimated from SED fitting with those predicted using the main model for training data (left) and validation data (right) sets. The orange line presents one-to-one correlation and the red dashed-line indicates the best fit. {The equation of the best-fit is displayed on each plot.}}
  \label{fig:main_model}
\end{figure*}

In Figure \ref{fig:main_model}, we compare the true IR luminosities estimated from SED fitting with those predicted using the main feature set. {The close alignment of the one-to-one correlation (yellow line) with the best-fit (red dashed line), with slopes of 0.96 for the training data and 0.94 for the validation data, indicates a strong linear relationship between the variables.} This alignment is supported by small RMSE values and no offset, indicating precise predictions across the dataset. However, minor deviations are observed where predictions slightly exceed true values at lower IR luminosities and slightly underestimate them at higher luminosities. Despite these minor deviations, our network estimates IR luminosity with uncertainties of 0.22 dex for the training data and 0.26 dex for the validation data, both at a 95\% confidence level{, considering the normal distribution of residuals (predicted values – true values) for both training and validation data.} This demonstrates that the model has successfully learned the complex relationships between input features and IR luminosity, leading to good predictions across both training and validation sets.

\section{Discussion}\label{discussion}
In this section, we explore how well our network generalizes to unseen data that the model has not encountered during the training process, ensuring it extends beyond the training set. {We convert the predicted IR luminosities obtained from our main model (see Table 2) into star formation rates ($\rm SFR_{ANN}$) using the modified equation for a Chabrier IMF:}

\begin{equation}
    \rm log\ SFR(M_{\odot} yr^{-1})\ =\ log\ L_{IR}(erg\ s^{-1}) -43.591
    \label{eq:LIR_SFR}
\end{equation}

{We validate our model by comparing its predictions of SFRs with those from three specific catalogs: the MPA-JHU value-added catalog based on DR8, the GALEX-SDSS-WISE Legacy Catalog \citep[GSWLC-2,][]{Salim16, Salim18}, and the catalog provided by  \cite{Ellison16}. These catalogs were chosen for their comprehensive size and because they are primarily based on SDSS data. Importantly, all input features used in our main network are available for the majority of targets in these catalogs, enabling us to apply our model to them. 
 Additionally, each catalog provides independent SFR estimations, ensuring that the performance of our model can be evaluated against other SFR indicators. By evaluating our model on these datasets, we assess its reliability in predicting SFRs for new, unseen data, highlighting its potential for broader astrophysical research. In the following, we describe these catalogs in detail, explain the construction of our unseen sample, and discuss our results.}

{First, the MPA-JHU catalog offers SFR measurements based on $\rm H\alpha$ emission line flux corrected for dust extinction that can obscure star formation activity using the Balmer decrement for galaxies classified as star-forming according to the BPT diagram \citep[]{Baldwin81}. For AGNs or galaxies lacking significant emission lines, SFRs are alternatively derived from the relation between the 4000 \AA~break strength ($\rm D_{n}4000$ \AA) and SFR, following methods such as those outlined by \cite{Kauffmann03}. Additionally, all SFR measures in the MPA-JHU catalog are adjusted for fiber aperture effects, which is crucial for providing consistent SFR estimates across different parts of the galaxies. These SFR estimates initially assume a Kroupa IMF but are then recalibrated to a Chabrier IMF, ensuring compatibility with other commonly used stellar population models. We focus on targets classified as star-forming, composite, and AGN in this catalog.} 

{Second, we utilize the GSWLC-2 catalog, the second version of GSWLC that provides SFR values obtained from UV/optical SED fitting and mid-IR SFRs derived from IR templates based upon 22 $\rm \mu m$ WISE observations. As these mid-IR SFRs tend to be overestimated for galaxies with AGN, likely due to non-stellar dust heating \citep[][]{Salim16}, we adopt the UV/optical SFRs for our comparisons. We exclude targets with flagged SEDs in the GSWLC-2 catalog.} 

{Third, we utilize a catalog containing predicted total IR luminosities for a substantial number of SDSS galaxies. These IR luminosities are estimated using an artificial neural network (ANN) that incorporates 23 input optical parameters, including continuum and emission line fluxes \citep[]{Ellison16}. These estimates are the most relevant for comparison with our network. we convert the total IR luminosity into SFR using Equation \ref{eq:LIR_SFR}.} We prioritize targets with more reliable SFR estimates (i.e., $\rm \sigma_{ANN} < 0.1$) as suggested by \cite{Ellison16}. 

We note that for each catalog, we collect and clean the data using the same process applied to the input data. {To ensure a fair comparison of our learning model across these catalogs, we cross-correlate the samples before applying the model, resulting in the selection of the same individual objects and an equal number of targets across the datasets.} This involves aligning and standardizing the data to eliminate inconsistencies and ensuring that each sample is directly comparable under the same conditions. This yields $\sim$ 45k targets.

\begin{figure*}[]
    \centering
    \subfigure{\includegraphics[width=0.84\textwidth]{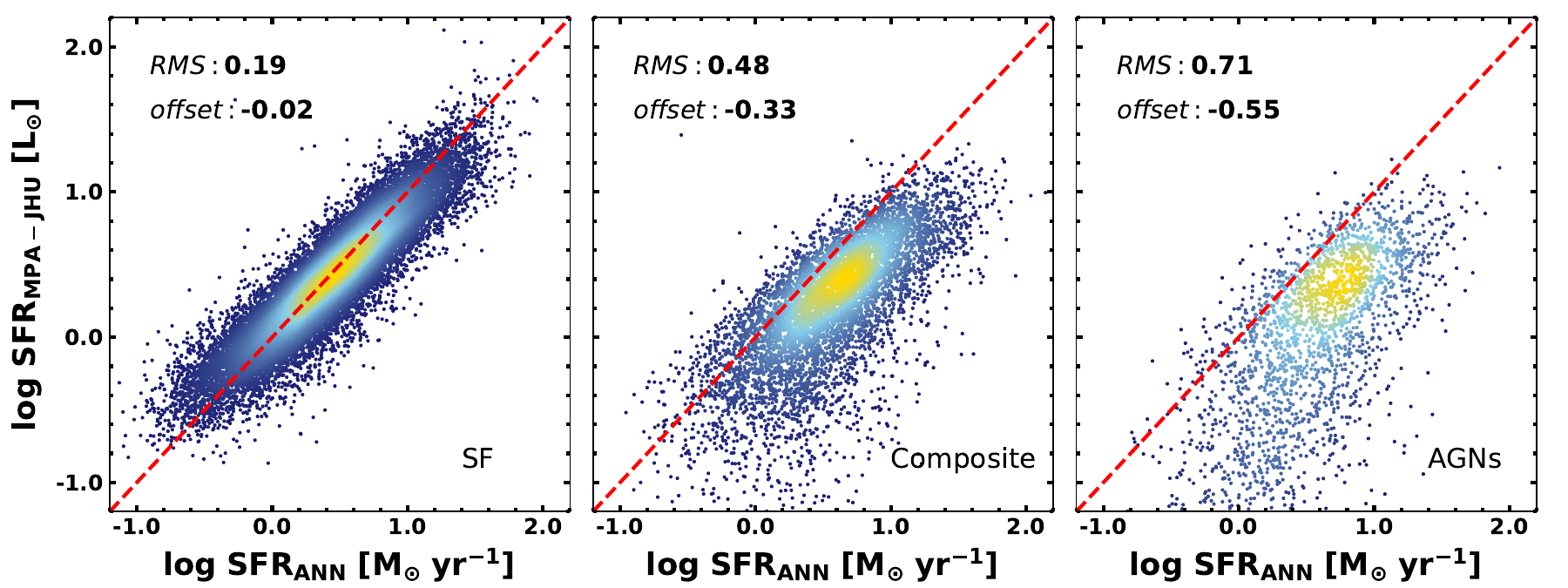}} \\
    \subfigure{\includegraphics[width=0.84\textwidth]{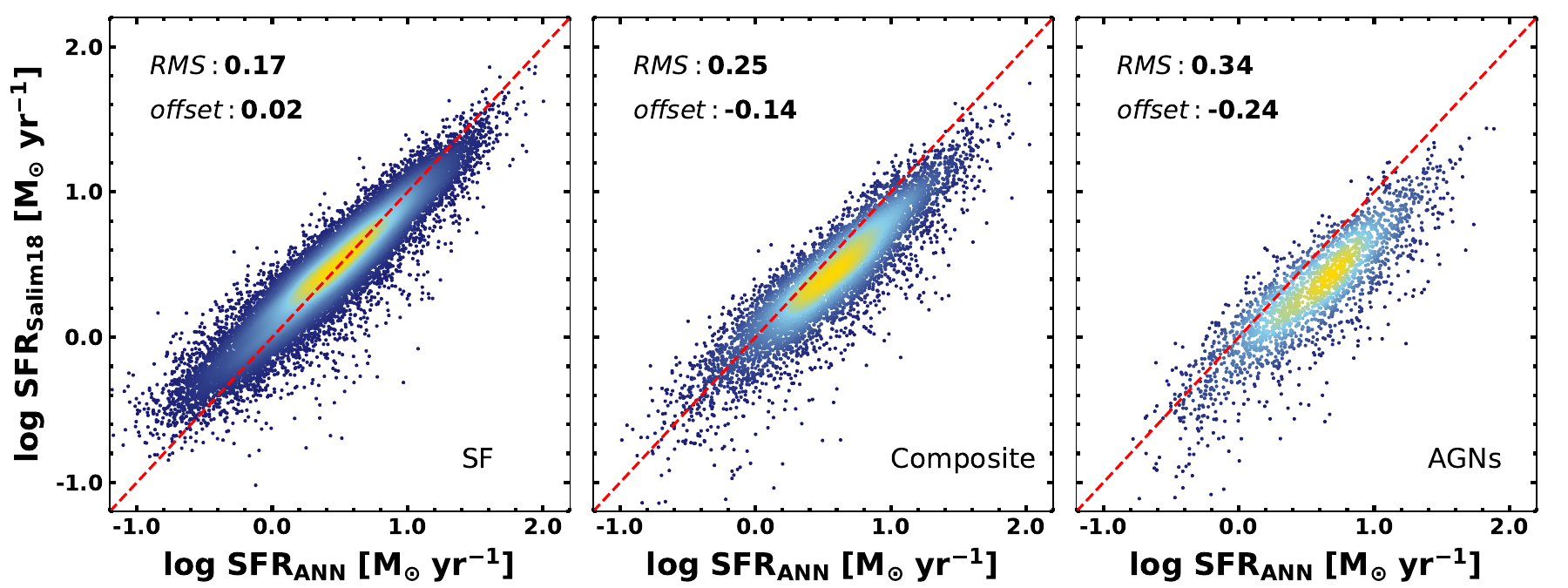}} \\
    \subfigure{\includegraphics[width=0.84\textwidth]{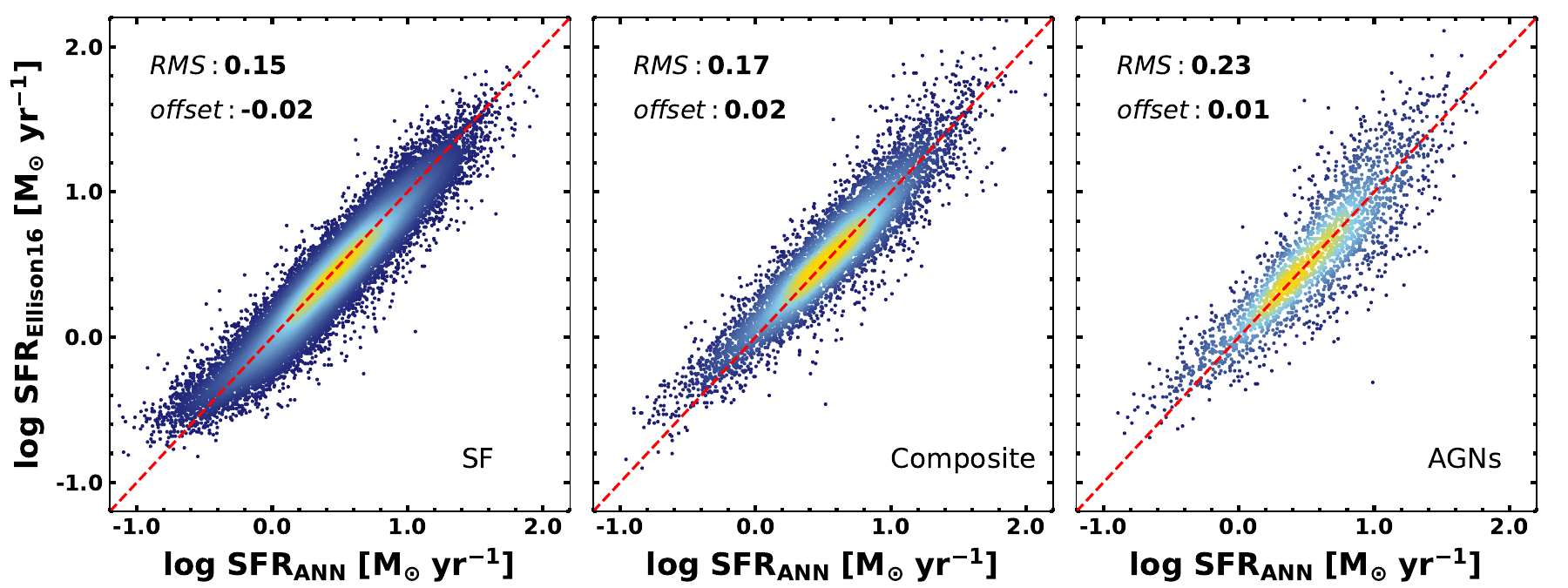}}
    \caption{The comparison between our ANN results (X-axis) with MPA-JHU catalog (top row), GSWLC-2 catalog (middle row) and ANN trained by \cite{Ellison16} (bottom row) for star-forming galaxies, composites and AGNs.}
    \label{fig:compared_catalogs}
\end{figure*}

Likewise, in Figure \ref{fig:compared_catalogs}, we compare our network output with SFRs from these catalogs. 
{For this comparison, we divide the sample into star-forming galaxies, composite galaxies, and AGNs to explore any systematic differences arising from AGN activity, which can influence SFR measurements due to contamination from non-stellar emission. The top row compares our network predictions and the MPA-JHU catalog. For star-forming galaxies (left panel), there is an excellent agreement between the predicted and observed SFRs, with an RMSE $\sim$ 0.2 dex and offset = 0.02 dex. This result underscores the good performance of our network in accurately predicting SFRs for galaxies dominated by star-forming processes. In our training data, star-forming galaxies are the most represented, followed by composite galaxies, with AGNs having the fewest targets. This imbalance likely contributes to the better performance of ANN for star-forming galaxies. Additionally, the well-modeled IR SEDs of star-forming galaxies provide more reliable true SFR values for training the model. Together, these factors enhance the ability of our network to generalize and predict accurately for this galaxy population.} 

{However, the agreement is less robust for composite galaxies (middle panel), with an RMSE $\sim$ 0.5 dex and an offset of 0.33 dex. These discrepancies might stem from the dual contributions of star formation and AGN activity in composite galaxies, which complicate the determination of accurate SFRs.} 

{For AGNs (right panel), the RMSE increases to $\sim$ 0.7 dex, with a significant offset of 0.55 dex. The larger scatter and systematic offset may arise from the considerable uncertainties associated with using  $\rm D_{n}4000$-based SFR estimates for AGN-hosting galaxies, as AGN activity can dominate or obscure the spectral features tied to star formation. Alternatively, these discrepancies could point to limitations in the ability of our ANN to predict IR luminosities accurately in galaxies with prominent AGN activity, potentially due to insufficient training data or inherent complexities in the underlying physics of AGN-dominated systems.} 

In the middle row, we present the comparisons between our neural network and the GSWLC-2 catalog. For star-forming galaxies (left panel), there is good agreement, with RMSE $\sim0.2$ and offset = 0.02 dex, indicating consistent performance. However, for composite galaxies (middle panel), there are still some discrepancies, reflected by an RMSE $\sim 0.25$ and offset = 0.14 dex. Despite some improvements, noticeable offsets persist for AGNs (right panel), with an RMSE $\sim 0.3$ and an offset of 0.24 dex, {suggesting that further refinement may be required}.

The bottom row depicts the comparison between the output of our network and the SFR predicted by the ANN trained by \cite{Ellison16}. We observe excellent agreement between the two networks for all galaxies. However, the superiority of our network lies in its ability to be applied to a larger population of galaxies. This is because we trained our network without relying on emission line measurements, which require reliable spectra.

{Overall, our neural network demonstrates good performance, particularly for star-forming galaxies, where it consistently predicts SFRs with minimal discrepancies compared to established catalogs. While some offsets persist for composite galaxies and AGNs. These discrepancies might be attributed to the lack of accurate estimations in the catalogs or insufficient training of our network due to the small number of composite galaxies and AGNs in the training data or because the IR-SEDs for these types may not be well-modeled. This issue could be addressed in future developments by including larger, more balanced datasets with better class representation, which would enhance the ability of the trained model to predict SFRs across all galaxy types. The current study suggests that further refinement may be required but also emphasizes the robustness of our model, particularly in its ability to generalize across different galaxy populations.}

\section{Summary}\label{summary}

{In this study, we develop and train an artificial neural network using a dataset of FIR-detected galaxies to accurately predict their IR luminosities. We explore five distinct sets of input features, including optical and MIR photometric data, SDSS redshift, and their corresponding errors. Subsequently, we incorporate additional features, such as NIR and UV photometry, as well as emission lines, in a stepwise manner. Each time, we retrain the network with the augmented data (see Table \ref{tab:features}).}

{Our findings reveal that all sets of features yield comparable predictions of IR luminosity. Interestingly, incorporating all available photometric data led to an improvement in the predictive performance of our network, particularly in the case of star-forming galaxies. This suggests that integrating diverse photometric measurements enhances the ability of the model to capture the complex relationships between different spectral regions and the IR luminosities of galaxies.}

While the performance of our network closely resembles that of the previous study by \cite{Ellison16}, our distinct approach employing solely photometric data enables its application to a broader range of galaxies, enhancing its utility across a large population of galactic systems.

This work has been supported by the Basic Science Research Program through the National Research Foundation of the Korean Government
(grant No. 2021R1A2C3008486) and Institute of Information \& communications Technology Planning \& Evaluation (IITP) grant funded by the Korea government(MSIT) (No.RS-2021-II212068, Artificial Intelligence Innovation Hub).\\

{Software: CIGALE \citep[]{Boquien19, Yang20}, Astropy \citep[]{Astropy22, Astropy18, Astropy13}, Tensorflow \citep[]{Abadi15},  Scikit-learn \citep[]{Pedregosa18}}\\

{Data availability: The predicted values of SFR for unseen data are publicly available online.}

\bibliography{ref}
\end{document}